\documentclass[utf8]{frontiersSCNS} % for Science, Engineering and Humanities and Social Sciences articles
\usepackage{url,hyperref,lineno,microtype,subcaption}
\usepackage[onehalfspacing]{setspace}

\usepackage{amsmath, amssymb, bbm}
\usepackage{graphicx}
\usepackage{fancybox}
\usepackage{cite}
\usepackage{mathrsfs}
\usepackage[usenames,dvipsnames]{xcolor}
\usepackage{booktabs,longtable}
\usepackage{placeins}

\newcommand{\de}{\mathrm{d}}

\renewcommand{\vec}[1]{\boldsymbol #1}

\newcommand{\h}{\hspace{1pt}}
\newcommand{\mh}{\hspace{-1pt}}
\newcommand{\hh}{\hspace{0.5pt}}
\newcommand{\e}{\mathrm e}

\DeclareOldFontCommand{\rm}{\normalfont\rmfamily}{\mathrm}

\DeclareMathAlphabet{\mathbbmsl}{U}{bbm}{m}{sl}

\usepackage[T1]{fontenc}

\def\keyFont{\fontsize{8}{11}\helveticabold }
\def\firstAuthorLast{Schober {et~al.}} %use et al only if is more than 1 author
\def\Authors{Giulio A.~H.~Schober\,$^{1,*}$, Jannis Ehrlich\,$^{1,2}$, Timo Reckling\,$^{1}$, and Carsten Honerkamp\,$^{1,3}$}
\def\Address{$^{1}$Institute for Theoretical Solid State Physics, RWTH Aachen University, Aachen, Germany 
\\ $^{2}$ Peter Gr\"unberg Institute and Institute for Advanced Simulation, Forschungszentrum J\"ulich, Germany \\
$^{3}$ JARA-FIT, J\"ulich Aachen Research Alliance - Fundamentals of Future Information Technology, Germany }
\def\corrAuthor{Institute for Theoretical Solid State Physics, RWTH Aachen University, Otto-Blumenthal-Stra\ss e, 52074 Aachen, Germany}

\def\corrEmail{schober@physik.rwth-aachen.de}

\begin{document}
\onecolumn
\firstpage{1}

\title[TU-fRG for multiband systems with spin-orbit coupling]{Truncated-unity functional renormalization group for multiband systems with spin-orbit coupling} 

\author[\firstAuthorLast ]{\Authors} %This field will be automatically populated
\address{} %This field will be automatically populated
\correspondance{} %This field will be automatically populated

\extraAuth{}% If there are more than 1 corresponding author, comment this line and uncomment the next one.
%\extraAuth{corresponding Author2 \\ Laboratory X2, Institute X2, Department X2, Organization X2, Street X2, City X2 , State XX2 (only USA, Canada and Australia), Zip Code2, X2 Country X2, email2@uni2.edu}

\maketitle

\begin{abstract}

Although the functional renormalization group (fRG) is by now a well-established method for investigating correlated electron systems, it is still undergoing significant technical and conceptual improvements. In particular, the motivation to optimally exploit the parallelism of modern computing platforms has recently led to the development of the ``truncated-unity'' functional renormalization group (TU-fRG). Here, we review this fRG variant, and we provide its extension to multiband systems with spin-orbit coupling. Furthermore, we discuss some aspects of the implementation and outline  opportunities and challenges ahead for predicting the ground-state ordering and emergent energy scales for a wide class of quantum materials.
%%% Leave the Abstract empty if your article does not require one, please see the Summary Table for full details.

\tiny
 \keyFont{ \section{Keywords:} functional renormalization group, interacting fermions, high-performance computing, multiband systems, spin-orbit coupling, quantum materials} %All article types: you may provide up to 8 keywords; at least 5 are mandatory.
\end{abstract}

\section{Introduction}

The rapid progress of condensed matter physics in recent decades, in its full breadth from fundamental science to technological applications, has mainly been based on the systematic development and strengthening of two main pillars: {\itshape materials} and {\itshape methods.}

On the one hand, the discovery of  new classes of {\itshape quantum materials} such as high-temperature superconductors and topological insulators has brought forth a rich variety of new effects and astonishing phenomena (for recent reviews see e.g.~\citep{Mackenzie17, Sun17, Xu17}). Among others, spin-orbit coupling has been identified as an important player in the formation of exotic phases with unconventional bulk and surface properties \citep{Winkler,Smidman17}. Consequently, {\itshape multiband systems with spin-orbit coupling} (for some references, see e.g.~\citep{Schober}) are by now a major topic of cutting-edge research in condensed matter physics, which largely overlaps with the fields of spintronics and topological matter. Their study has led to new theoretical insights regarding, for example, relativistic effects in solids \citep{Lee, Laszlo12} and emerging entities such as Majorana quasiparticles \citep{Lutchyn10, Oreg10}. In addition, quantum materials with spin-orbit coupling hold the promise for technological applications ranging from high-performance, high-density memories to fault-tolerant topological information processing \citep{Sinova,Wu17}. More concretely, the Rashba model has become a paradigm for coupling the spin and orbital degrees of freedom \citep{Rashba12}. While having traditionally been observed at surfaces or interfaces between different materials, more recently, BiTeX (X = Cl, Br and I) as well as GeTe have attracted much interest as giant {\itshape bulk} Rashba semiconductors \citep{BiTeIMagn, Schwalbe16, Saeed17, Picozzi14, Liebmann16}. Hence, these materials are also promising candidates for spintronics applications such as the Datta-Das spin transistor \citep{Datta89, Koo09}. In this context, we also mention the recent proposal \citep{Ciftja} of a spintronic device which does not require the injection of spin-polarized electrons from one quantum dot to the other.

On the other hand, exotic material behavior often originates from electronic correlations, and the quest for new computational methods capable of dealing with these correlations is a driving force in modern materials physics. While the most prominent example for a non-trivial interaction-induced quantum phase may be an ordinary superconductor, more recently, interaction-driven topological phases have also attracted much interest  (see e.g.~\citep{Dzero12,Isobe15}). In particular, the {\itshape interplay between spin-orbit coupling and electron-electron interactions} gives rise to a plethora of unconventional quantum effects such as mixed singlet-triplet superconductivity induced by the Rashba coupling \citep{GR01,Schober,Vijay16}. However, simultaneously treating the enormous number of electrons in a solid requires the development of new computational approaches, which on the one hand capture the essential physics and on the other hand allow for practical calculations on reasonable time scales. Fortunately, the recent advancement of condensed matter science has come along with an equally steep increase in the available computational resources, which is mainly due to the evolution of building blocks of large computing architectures from single-core CPUs to compute nodes with multiple cores. This development naturally requires the adaption of modern calculation methods to optimally exploit the parallelism of modern computing platforms.

In this article, we summarize recent progress in the field of fRG for fermions in solids, which was achieved by the development of the {\itshape truncated-unity functional renormalization group} (TU-fRG) \citep{Julian17, Julian17Book, David17}. Moreover, we take one step forward by adapting this method for application to multiband systems with spin-orbit coupling.

To begin with, the functional renormalization group (fRG) is a field-theoretical approach to the electronic many-body problem, which is capable of treating the different energy scales in the electronic spectrum of a solid as well as the different---and possibly competing---ordering tendencies at low temperatures \citep{Kopietz,Metzner,Platt}. Having undergone various phases of exploration and refinement, it is by now regarded as an unbiased method with the potential of reaching quantitative precision in the prediction of energy scales and parameter ranges \citep{Julian17}. The different energy scales are treated by successively ``integrating out'' the high-energy degrees of freedom in the path integral formalism and thereby deriving effective interactions for the low-energy degrees of freedom. In a solid, the latter correspond to the electrons at the Fermi level, hence the fRG can predict the ground state ordering of the many-electron system. As compared to exact methods such as lattice quantum Monte Carlo (QMC) simulations, the fRG has the advantage of not being limited strongly in the choice of tight-binding and interaction parameters (which is necessary in the case of QMC to avoid the occurrence of a sign problem) \citep{David17Coul}. Until recently, the fRG was usually combined with a Fermi surface patching approximation, whereby it has been successfully applied to various material classes such as high-temperature superconductors \citep{Giering, Eberlein14, Lichtenstein}, mono- and few-layer graphene \citep{Classen, Lang, Pena, Scherer2}, and systems with spin-orbit coupling \citep{Maier, Dscherer, Schober}. For recent reviews of the fRG in solid-state physics, we refer the interested reader to Refs.~\citep{Metzner, Platt, Schober}.

As a recent variant of the fRG technique, the TU-fRG has been developed and applied to extended Hubbard models on the square and honeycomb lattices \citep{Julian17, Julian17Book,David17,David17Coul}. This method is based on earlier approaches called the {\itshape exchange parametrization fRG} \citep{Husemann09} (see also \citep{Husemann12, Eberlein10, Maier13}) and the {\itshape singular-mode fRG} (SM-fRG) \citep{Wang12}. As compared to the exchange parametrization fRG for the two-fermion interaction, the TU-fRG introduces additional insertions of truncated partitions of unity, which decouple the fermionic propagators from the exchange propagators. This leads to a separation of the underlying differential equations and therefore enables an efficient parallelization on a large number of compute nodes (for details on the numerical implementation, see \citep{Julian17, David17, David17Coul}). Remarkably, in the TU-fRG implementation, the CPU time scales only linearly with the momentum resolution of the two-fermion vertex as compared to a quartic dependence in the Fermi surface patching approximation. Therefore, the TU-fRG allows one to capture the wavevector dependence of the effective two-fermion vertices with an unrivaled precision \citep{Julian17}, thus paving the way for quantitative predictions of phase diagrams, leading correlations and emergent energy scales in multiband systems with realistic {\itshape ab initio} interaction parameters. This procedure has already been demonstrated in Ref.~\citep{David17} for the case of graphene. Most recently, it has even been applied in parameter ranges which are not accessible to QMC simulations \citep{David17Coul}. 

So far, the application of the TU-fRG has been restricted to spin-SU(2)-invariant systems. In fact, this symmetry is exploited in the very derivation of the flow equations \citep{Julian17} (see also \citep{SH00}). Since this restriction excludes the important class of materials with spin-orbit coupling, it is highly desirable to derive a generalization of the TU-fRG which does not assume the SU(2) symmetry of the single-particle Hamiltonian. In this generalization, the two-fermion vertices will not only acquire additional spin indices, but also the general structure of the flow equations will be modified. Apart from providing a general perspective on the TU-fRG technique, it is precisely the goal of the present article to derive such generalized flow equations. In order to keep the formalism simpler we will not discuss the treatment of the frequency dependence of the interactions (for a recent treatment, see \citep{Eberlein15,Vilardi17}). Furthermore, we will not consider self-energy corrections on the internal lines (for some discussion of this aspect, see \citep{Metzner,Eberlein15,Vilardi17}). We remark that in the SM-fRG context, generalized flow equations for non-SU(2)-invariant systems have already been derived \citep{Wang4} and successfully applied to various materials with spin-orbit coupling \citep{Wang1,Wang3,Wang2}.

This article is organized as follows: In \S\,\ref{sec_channel}, we define the projections of an arbitrary two-particle vertex onto the particle-particle, crossed particle-hole and direct particle-hole channels. In \S\,\ref{sec_rge}, we first define the exchange propagators from the respective single-channel coupling functions. Subsequently, we derive the flow equations for these exchange propagators, and we show that the number of relevant equations can be reduced by using an antisymmetry relation between crossed and direct particle-hole terms. Finally, the set of TU-fRG equations is completed by the cross projections between the particle-particle and particle-hole terms, for which we also provide explicit expressions. In \S\,\ref{sec_orb_band}, we derive the transformation laws for the exchange propagators between the orbital (or spin) and the band basis, and we briefly compare these two representations with respect to the numerical effort required to solve the flow equations. In \S\,\ref{sec_appl}, we benchmark the generalized TU-fRG equations by applying them to the Rashba model with a local interaction. Restricting us to the particle-particle channel, we provide an analytic solution which is consistent with the ladder resummation of Ref.~\citep{Schober}. Finally, we give an outlook on future applications of the TU-fRG to realistic multiband models with spin-orbit coupling.

\section{Channel decomposition} \label{sec_channel}

We put our main focus on the fRG for the two-particle vertex $V$ (see e.g.~\citep{SH00}). The renormalization group equation (RGE) for this vertex is an ordinary differential equation with respect to a scaling parameter $\Lambda$, which has contributions from three different channels: {\itshape particle-particle, crossed particle-hole, and direct particle-hole channel} (see \S\,\ref{sec_rge}). An analysis as well as previous numerical investigations reveal that each channel has a strong dependence on one particular momentum contribution \citep{Metzner}. Therefore, we define the functionals $\hat P[V]$, $\hat C[V]$ and $\hat D[V]$ as projections onto these channels, which are parametrized by their respective main momentum such that the following equations hold (we use similar conventions as in Ref.~\citep{Eckhardt}):
\begin{align}
 V(\vec k_1, \vec k_2, \vec k_3) & = \sum_{\vec R, \h \vec R'} \hat P[V]_{\vec R \vec R'}(\vec k_1 + \vec k_2) \, f_{\vec R}^*(\vec k_1) \h f_{\vec R'}(\vec k_4) \,, \label{P_1} \\[5pt]
 V(\vec k_1, \vec k_2, \vec k_3) & = \sum_{\vec R, \h \vec R'} \hat C[V]_{\vec R \vec R'}(\vec k_1 - \vec k_3) \, f_{\vec R}^*(\vec k_1) \h f_{\vec R'}(\vec k_4) \,, \label{C_1} \\[5pt]
 V(\vec k_1, \vec k_2, \vec k_3) & = \sum_{\vec R, \h \vec R'} \hat D[V]_{\vec R \vec R'}(\vec k_3 - \vec k_2) \, f_{\vec R}^*(\vec k_1) \h f_{\vec R'}(\vec k_3) \,. \label{D_1}
\end{align}
Here, $\vec k_1$ and $\vec k_2$ are ingoing momenta, while $\vec k_3$ and $\vec k_4 \equiv \vec k_1 + \vec k_2 - \vec k_3$ are outgoing momenta of the vertex function (where momentum conservation follows from translational invariance). Furthermore, the {\itshape form factors} $f_{\vec R}(\vec k)$ are labeled by an index $\vec R$, which may correspond to the (Bravais) lattice coordinates in the case of plane-wave functions, $f_{\vec R}(\vec k) = \exp(-\textnormal i \vec k \cdot \vec R)$. On the real lattice, the form factors then become bond selectors, $f_{\vec R}(\vec R') = \delta_{\vec{R},\hh\vec{R}'}$. This provides a very natural description of the interaction as the sum of interacting fermion bilinears of particle-particle ($P$) or particle-hole type ($C$,\,$D$) that live on bonds of length $\vec{R}$. In such a picture, the truncation of the sums in Eqs.~\eqref{P_1}, \eqref{C_1}, and \eqref{D_1} beyond a certain $|\vec{R}_{\rm c}|$ just means that interactions of bilinears with a range longer than   $|\vec{R}_{\rm c}|$ are ignored.  Note that this truncation in the pair length does not imply a restriction on the length of the pairwise interaction between the bilinears kept. Depending on the symmetry of the problem under consideration, it may be useful to form linear combinations of the form factors that transform according to a specific irreducible representation of the point group, as done in Ref.~\citep{Julian17} or explained in Ref.~\citep{Platt}.
More generally, we only assume that the form factors constitute an orthonormal basis in the following sense:
\begin{align}
 \frac 1 N \sum_{\vec k} f_{\vec R}(\vec k) \h f_{\vec R'}^*(\vec k) & = \delta_{\vec R, \hh \vec R'} \,, \\[5pt]
 \sum_{\vec R} f_{\vec R}(\vec k) \h f_{\vec R}^*(\vec k') & = N \h \delta_{\vec k, \hh \vec k'} \,,
\end{align}
where $N$ denotes the total particle number. We relabel the main momentum transfers as $\vec s = \vec k_1 + \vec k_2$, $\vec u = \vec k_1 - \vec k_3$, and $\vec t = \vec k_3 - \vec k_2$, such that Eqs.~\eqref{P_1}--\eqref{D_1} are equivalent to
\begin{align}
 V(\vec k, \h \vec s - \vec k, \h \vec s - \vec k', \h \vec k') & = \sum_{\vec R, \h \vec R'} f_{\vec R}^*(\vec k) \h f_{\vec R'}(\vec k') \, \hat P[V]_{\vec R \vec R'}(\vec s) \,, \label{def_P_inv} \\[5pt]
 V(\vec k, \h \vec k' - \vec u, \h \vec k - \vec u, \h \vec k') & = \sum_{\vec R, \h \vec R'} f_{\vec R}^*(\vec k) \h f_{\vec R'}(\vec k') \, \hat C[V]_{\vec R \vec R'}(\vec u) \,, \\[5pt]
 V(\vec k, \h \vec k' - \vec t, \h \vec k', \h \vec k - \vec t) & = \sum_{\vec R, \h \vec R'} f_{\vec R}^*(\vec k) \h f_{\vec R'}(\vec k') \, \hat D[V]_{\vec R \vec R'}(\vec t) \,,
\end{align}
where, for the sake of clarity, we have written out explicitly the fourth momentum argument of the vertex function (which is determined by momentum conservation).
The converse equations, which can be regarded as explicit definitions of the functionals (projections onto the different channels) $\hat P[V]$, $\hat C[V]$ and $\hat D[V]$, then read as follows:
\begin{align}
 \hat P[V]_{\vec R \vec R'}(\vec s) & = \frac{1}{N^2} \sum_{\vec k, \h \vec k'} f_{\vec R}(\vec k) \h f^*_{\vec R'}(\vec k') \, V(\vec k, \h \vec s - \vec k, \h \vec s - \vec k', \h \vec k') \,, \label{def_P} \\[5pt]
 \hat C[V]_{\vec R \vec R'}(\vec u) & = \frac{1}{N^2} \sum_{\vec k, \h \vec k'} f_{\vec R}(\vec k) \h f^*_{\vec R'}(\vec k') \, V(\vec k, \h \vec k' - \vec u, \h \vec k - \vec u, \h \vec k') \,, \\[5pt]
 \hat D[V]_{\vec R \vec R'}(\vec t) & = \frac{1}{N^2} \sum_{\vec k, \h \vec k'} f_{\vec R}(\vec k) \h f^*_{\vec R'}(\vec k') \, V(\vec k, \h \vec k' - \vec t, \h \vec k', \h \vec k - \vec t) \,.
\end{align}
The conventions that we have chosen here have the following advantages (compare Refs.~\citep{Wang12, Julian17}):
\begin{enumerate}
 \item Each form factor in Eqs.~\eqref{P_1}--\eqref{D_1} depends on only one momentum, which is part of the original vertex. This will facilitate the following derivations. \smallskip
 \item If two vertices fulfill $V(\vec k_1, \vec k_2, \vec k_3, \vec k_4) = -W(\vec k_1, \vec k_2, \vec k_4, \vec k_3)$, then
 \begin{equation} \hat D[V]_{\vec R \vec R'}(\vec t) = -\hat C[W]_{\vec R \vec R'}(\vec t) \,. \label{DCsym} \end{equation} In particular, if the vertex $V$ is antisymmetric with respect to its outgoing momenta such that $V(\vec k_1, \vec k_2, \vec k_3, \vec k_4) = -V(\vec k_1, \vec k_2, \vec k_4, \vec k_3)$, then
 \begin{equation}
 \hat D[V] = -\hat C[V] \,.
 \end{equation}
 These identities will become crucial in \S\,\ref{subsec_antisym}. \smallskip
  \item If $V$ is hermitian in the sense that $V(\vec k_1, \vec k_2, \vec k_3, \vec k_4) = V^*(\vec k_4, \vec k_3, \vec k_2, \vec k_1)$, then
 \begin{equation}
  \hat P[V]_{\vec R \vec R'}(\vec s) = \hat P[V]^*_{\vec R' \vec R}(\vec s) \,,
 \end{equation}
i.e., the matrix $\hat P[V](\vec s)$ is also hermitian for any $\vec s$. The same applies to the matrix $\hat C[V](\vec u)$, and to $\hat D[V](\vec t)$ provided that $V(\vec k_1, \vec k_2, \vec k_3, \vec k_4) = V(\vec k_2, \vec k_1, \vec k_4, \vec k_3)$.
\end{enumerate}
Finally, we remark that by definition all wavevectors including $\vec s, \vec u, \vec t$ \h are restricted to the first Brillouin zone. For example, Eq.~\eqref{P_1} is a shorthand notation for
\begin{equation}
 V(\vec k_1, \vec k_2, \vec k_3) = \sum_{\vec R, \h \vec R'} f_{\vec R}^*(\vec k_1) \h f_{\vec R'}(\vec k_4) \, \sum_{\vec s} \sum_{\vec K} \delta_{\vec s + \vec K, \h \vec k_1 + \vec k_2} \, \hat P[V]_{\vec R \vec R'}(\vec s) \,.
\end{equation}
Here, we formally sum over all reciprocal lattice vectors $\vec K$, but the condition that $\vec k_1$, $\vec k_2$, and $\vec s$ all lie in the first Brillouin zone fixes precisely one vector $\vec K$ which gives a non-vanishing contribution.

\section{Renormalization group equations} \label{sec_rge}

\subsection{Derivation} \label{subsec_deriv}

We now consider a general multiband system, where the vertex additionally depends on four {\itshape band indices}, which we denote by Latin letters. (By contrast, orbital or spin indices will be denoted by Greek letters, see \S\,\ref{sec_orb_band}.) In the general case without SU(2) symmetry, the RGE in the band basis reads as follows (see \citep{SH00}, or \citep[Eqs.~(82)--(89)]{Schober}):
\begin{equation} \label{rg_eq}
 \frac{\de}{\de \Lambda} (V_\Lambda)_{n_1 n_2 n_3 n_4}(\vec p_1, \h \vec p_2, \h \vec p_3)
 = \big[ \varPhi_{\Lambda}^{\rm pp} + \h \varPhi_{\Lambda}^{\rm ph,c} + \h \varPhi_{\Lambda}^{\rm ph,d} \h \big]_{n_1 n_2 n_3 n_4} (\vec p_1, \h \vec p_2, \h \vec p_3) \,,
\end{equation}
where the particle-particle, crossed particle-hole and direct particle-hole terms are given by (suppressing $\Lambda$~dependencies in the notation)
\begin{align}
 & \varPhi^{\rm pp}_{n_1 n_2 n_3 n_4}(\vec p_1, \h \vec p_2, \h \vec p_3) = \label{eq_phipp} \\[3pt] \nonumber
 & -\frac 1 N \sum_{\ell_1, \h \ell_2} \sum_{\vec k} V_{n_1 n_2 \ell_1 \ell_2}(\vec p_1, \h \vec p_2, \h \vec k) \, L^-_{\ell_1 \ell_2}(\vec k, \h \vec p_1 + \vec p_2 - \vec k) \, V_{\ell_1 \ell_2 n_3 n_4}(\vec k, \h \vec p_1 + \vec p_2 - \vec k, \h \vec p_3) \,, \\[6pt]
 & \varPhi^{\rm ph,c}_{n_1 n_2 n_3 n_4}(\vec p_1, \h \vec p_2, \h \vec p_3) \label{eq_phiphd} = \\[3pt] \nonumber
 & \frac 2 N \sum_{\ell_1, \h \ell_2} \sum_{\vec k} V_{n_1 \ell_2 n_3 \ell_1}(\vec p_1, \h \vec k + \vec p_3 - \vec p_1, \h \vec p_3) \, L^+_{\ell_1 \ell_2}(\vec k, \h \vec k + \vec p_3 - \vec p_1) \, V_{\ell_1 n_2 \ell_2 n_4}(\vec k, \h \vec p_2, \h \vec k + \vec p_3 - \vec p_1) \,, \\[6pt]
 & \varPhi^{\rm ph,d}_{n_1 n_2 n_3 n_4}(\vec p_1, \h \vec p_2, \h \vec p_3) = -\varPhi^{\rm ph,c}_{n_1 n_2 n_4 n_3}(\vec p_1, \h \vec p_2, \h \vec p_1 + \vec p_2 - \vec p_3) \,. \label{eq_phiphc}
\end{align}
Furthermore, the {\itshape particle-particle loop} $L^-$ and the {\itshape particle-hole loop} $L^{+}$ are defined as
\begin{equation} \label{def_lmp}
 L^\mp_{\ell_1 \ell_2}(\vec k_1, \vec k_2)= \frac{\de}{\de\Lambda} \h \Big( \chi(e_{\ell_1}\mh(\vec k_1) ) \,\h \chi(e_{\ell_2}\mh(\vec k_2) ) \Big) \, F^\mp_{\ell_1 \ell_2}(\vec k_1, \vec k_2) \,,
\end{equation}
with the functions $F_\Lambda^{\mp}$ given by
\begin{align} \label{disc_2a}
 F^-_{\ell_1 \ell_2}(\vec k_1, \vec k_2) & = \frac{1 - f(e_{\ell_1}\mh(\vec k_1) ) - f(e_{\ell_2}\mh(\vec k_2) )}{e_{\ell_1}\mh(\vec k_1) + e_{\ell_2}\mh(\vec k_2)} \,, \\[5pt]  \label{disc_2b}
 F^+_{\ell_1 \ell_2}(\vec k_1, \vec k_2) & = \frac{f(e_{\ell_1}\mh(\vec k_1) ) - f(e_{\ell_2}\mh(\vec k_2) )}{e_{\ell_1}\mh(\vec k_1) - e_{\ell_2}\mh(\vec k_2)} \,.
\end{align}
Here, $e_\ell(\vec k) = E_\ell(\vec k) - \mu$ are the eigenvalues of the single-particle Hamiltonian measured relatively to the chemical potential, and $f(e) = (\e^{\beta e} + 1)^{-1}$ denotes the Fermi distribution function. We have assumed a momentum regulator $\chi_\Lambda(e(\vec k))$, which suppresses all momenta inside a shell of thickness $\Lambda$ around the Fermi surfaces \citep{Metzner}, but the above formulas can easily be generalized to other regulators (see e.g.~\citep{Husemann09}). Moreover, inherent in the above RGE is the level-two truncation, which neglects all Green functions with six or more external legs. Futher neglected are the self-energy and the frequency dependence of the four-point function. These approximations have already been applied successfully in many works before \citep{Metzner, Platt, Schober}.

We now define the scale-dependent {\itshape single-channel coupling functions} as follows (suppressing momentum dependencies to lighten the notation):
\begin{align}
 V^{(P)}_{n_1 n_2 n_3 n_4}(\Lambda) & := \int_{\Lambda_0}^\Lambda \de \Lambda' \, \varPhi^{\rm pp}_{n_1 n_2 n_3 n_4}(\Lambda') \,, \label{def_VP} \\[3pt]
 V^{(C)}_{n_1 n_2 n_3 n_4}(\Lambda) & := \int_{\Lambda_0}^\Lambda \de \Lambda' \, \varPhi^{\rm ph,c}_{n_1 n_2 n_3 n_4}(\Lambda') \,, \label{def_VD} \\[3pt]
 V^{(D)}_{n_1 n_2 n_3 n_4}(\Lambda) & := \int_{\Lambda_0}^\Lambda \de \Lambda' \, \varPhi^{\rm ph,d}_{n_1 n_2 n_3 n_4}(\Lambda') \,, \label{def_VC} 
\end{align}
such that the scale-dependent vertex can be decomposed into four contributions,
\begin{equation} \label{decom_V}
 V(\Lambda) = V^{(0)}  + V^{(P)}(\Lambda) + V^{(C)}(\Lambda) + V^{(D)}(\Lambda) \,,
\end{equation}
where $V^{(0)} \equiv V(\Lambda_0)$ is the initial interaction. Furthermore, we define the {\itshape exchange propagators} as the following matrices (sup\-{}pressing again $\Lambda$ dependencies):
\begin{align}
 P^{\vec R \vec R'}_{n_1 n_2 n_3 n_4}(\vec s) & := \hat P[V^{(P)}_{n_1 n_2 n_3 n_4}]_{\vec R \vec R'}(\vec s) \,, \label{def_mat_P} \\[5pt]
 C^{\vec R \vec R'}_{n_1 n_2 n_3 n_4}(\vec u) & := \hat C[V^{(C)}_{n_1 n_2 n_3 n_4}]_{\vec R \vec R'}(\vec u) \,, \label{def_mat_D} \\[5pt]
 D^{\vec R \vec R'}_{n_1 n_2 n_3 n_4}(\vec t) & := \hat D[V^{(D)}_{n_1 n_2 n_3 n_4}]_{\vec R \vec R'}(\vec t) \,. \label{def_mat_C}
\end{align}
Per constructionem, a scale derivative acting on these matrices yields the following RGE:
\begin{align}
 \dot P^{\vec R \vec R'}_{n_1 n_2 n_3 n_4}(\vec s) & = \hat P[\varPhi^{\rm pp}_{n_1 n_2 n_3 n_4}]_{\vec R \vec R'}(\vec s) \,, \label{rge_P} \\[5pt]
 \dot C^{\vec R \vec R'}_{n_1 n_2 n_3 n_4}(\vec u) & = \hat C[\varPhi^{\rm ph,c}_{n_1 n_2 n_3 n_4}]_{\vec R \vec R'}(\vec u) \,, \label{rge_D} \\[5pt]
 \dot D^{\vec R \vec R'}_{n_1 n_2 n_3 n_4}(\vec t) & = \hat D[\varPhi^{\rm ph,d}_{n_1 n_2 n_3 n_4}]_{\vec R \vec R'}(\vec t) \,. \label{rge_C}
\end{align}
To obtain a closed system of differential equations, one further has to project the right-hand sides onto the form-factor basis. For this purpose, we insert {\itshape partitions of unity} of the form-factor basis \citep{Julian17},
\begin{equation} \label{unity}
 1 = \sum_{\vec q} \delta_{\vec q, \hh \vec q'} = \frac 1 N \sum_{\vec q} \sum_{\vec R} f_{\vec R}^*(\vec q) \h f_{\vec R}(\vec q') \,,
\end{equation}
on both sides of the fermion loops $L^{\mp}$ in Eqs.~\eqref{eq_phipp}--\eqref{eq_phiphc}. After a lengthy but straightforward 
calculation (analogous to Ref.~\citep{Julian17}), we obtain the following RGE in matrix form, where we denote  by $\mathbbmsl P$, $\mathbbmsl C$ and $\mathbbmsl D$ the matrices with respective entries $P_{\vec R \vec R'}$, $C_{\vec R \vec R'}$, and $D_{\vec R \vec R'}$ (for the analogous equations in the SM-fRG framework, see \citep[Eqs.~(A3)]{Wang4}):
\begin{align}
 \dot{\mathbbmsl P}_{n_1 n_2 n_3 n_4}(\vec s) & = \sum_{\ell_1, \h \ell_2} \hat{\mathbbmsl P}[V_{n_1 n_2 \ell_1 \ell_2}](\vec s) \, \mathbbmsl L^-_{\ell_1 \ell_2}(\vec s) \, \hat{\mathbbmsl P}[V_{\ell_2 \ell_1 n_3 n_4}](\vec s) \,, \label{rge_Pmat} \\[5pt]
 \dot{\mathbbmsl C}_{n_1 n_2 n_3 n_4}(\vec u) & = 2 \h\sum_{\ell_1, \h \ell_2} \hat{\mathbbmsl C}[V_{n_1 \ell_2 n_3 \ell_1}](\vec u) \, \mathbbmsl L^+_{\ell_1 \ell_2}(\vec u) \, \hat{\mathbbmsl C}[V_{\ell_1 n_2 \ell_2 n_4}](\vec u) \,, \label{rge_Dmat} \\[5pt]
 \dot{\mathbbmsl D}_{n_1 n_2 n_3 n_4}(\vec t) & = -2 \h \sum_{\ell_1 \ell_2} \hat{\mathbbmsl D}[V_{n_1 \ell_2 \ell_1 n_4}](\vec t) \, \mathbbmsl L^+_{\ell_1 \ell_2}(\vec t) \, \hat{\mathbbmsl D}[V_{\ell_1 n_2 n_3 \ell_2}](\vec t) \,. \label{rge_Cmat}
\end{align}
Here, the loop matrices are given in terms of Eq.~\eqref{def_lmp} by
\begin{align}
 \big[ \mathbbmsl L^-_{\ell_1 \ell_2}(\vec s) \big]_{\vec R_1 \vec R_2} & = \frac 1 N \h \sum_{\vec q} L^-_{\ell_1 \ell_2}(\vec s - \vec q, \vec q) \, f_{\vec R_1}(\vec q) \h f_{\vec R_2}^*(\vec q) \,, \label{loopminus} \\[5pt]
 \big[ \mathbbmsl L^+_{\ell_1 \ell_2}(\vec u) \big]_{\vec R_1 \vec R_2} & = \frac 1 N \h \sum_{\vec q} L^+_{\ell_1 \ell_2}(\vec q, \h \vec q - \vec u) \, f_{\vec R_1}(\vec q) \h f_{\vec R_2}^*(\vec q) \,. \label{loopplus}
\end{align}
Finally, by substituting the decomposition \eqref{decom_V} into the right-hand sides of Eqs.~\eqref{rge_Pmat}--\eqref{rge_Cmat} and performing the projections in the various channels (see \S\,\ref{sec_proj}) we obtain a closed system of differential equations for the matrices $\mathbbmsl P$, $\mathbbmsl C$ and $\mathbbmsl D$. Since for implementing these flow equations numerically, the form-factor expansion in Eq.~\eqref{unity} has to be {\itshape truncated} appropriately (see the remarks in \S\,\ref{sec_channel}, and for a detailed discussion, see \citep{Julian17}), the resulting technique is called ``truncated-unity fRG'' or ``TUfRG''. 

\subsection{Antisymmetry of particle-hole terms} \label{subsec_antisym}

We now come to a crucial observation which allows us to further reduce the number of relevant channels in the generalized TU-fRG equations (where the SU(2) symmetry has not been exploited). In fact, the crossed and the direct particle-hole terms are directly related through Eq.~\eqref{eq_phiphc}, and  with 
the definitions \eqref{def_VD}--\eqref{def_VC}, this implies that
\begin{equation} \label{antisym_cd}
 V^{(D)}_{n_1 n_2 n_3 n_4}(\vec k_1, \vec k_2, \vec k_3) = -V^{(C)}_{n_1 n_2 n_4 n_3}(\vec k_1, \vec k_2, \vec k_1 + \vec k_2 - \vec k_3) \,.
\end{equation}
Further using the definitions \eqref{def_mat_D}--\eqref{def_mat_C} and the property \eqref{DCsym}, it follows that
\begin{equation}
 \mathbbmsl D_{n_1 n_2 n_3 n_4}(\vec t) = \hat{\mathbbmsl D}[V^{(D)}_{n_1 n_2 n_3 n_4}](\vec t) = -\hat{\mathbbmsl C}[V^{(C)}_{n_1 n_2 n_4 n_3}](\vec t) = -\mathbbmsl C_{n_1 n_2 n_4 n_3}(\vec t) \,. \label{antisym}
\end{equation}
Hence, the two matrices $\mathbbmsl C$ and $\mathbbmsl D$ are actually not independent of each other but simply related by the antisymmetry in the last two indices. As a consequence, one can show that the RGE \eqref{rge_Cmat} is in fact equivalent to Eq.~\eqref{rge_Dmat}, and can therefore be discarded. We are thus left with only two RGE for the matrices $\mathbbmsl P$ and $\mathbbmsl C$ as given by Eqs.~\eqref{rge_Pmat} and \eqref{rge_Dmat}. We remark that the antisymmetry property \eqref{antisym} is already well-established in SM-fRG works (see e.g.~\citep{Wang4}).

\subsection{Projections} \label{sec_proj}

It remains to perform the projections in Eqs.~\eqref{rge_Pmat}--\eqref{rge_Dmat} to obtain a closed system of RGE for the matrices $\mathbbmsl P$ and $\mathbbmsl C$. First, in the $P$-channel we have
\begin{equation}
 \hat{\mathbbmsl P}[V_{n_1 n_2 n_3 n_4}] = \hat{\mathbbmsl P}[V^{(0)}_{n_1 n_2 n_3 n_4}] + \mathbbmsl P_{n_1 n_2 n_3 n_4} + \hat{\mathbbmsl P}[V^{(C)}_{n_1 n_2 n_3 n_4}] + \hat{\mathbbmsl P}[V^{(D)}_{n_1 n_2 n_3 n_4}] \,, \label{sum_P}
\end{equation}
which follows directly from the decomposition \eqref{decom_V} and from the definition \eqref{def_mat_P}, i.e., $\mathbbmsl P = \hat{\mathbbmsl P}[V^{(P)}]$.
The second-last term can be evaluated as follows:
\begin{align}
 & \hat P[V^{(C)}_{n_1 n_2 n_3 n_4}]_{\vec R \vec R'}(\vec s)
 = \frac 1 {N^2} \h \sum_{\vec k, \h \vec k'} f_{\vec R}(\vec k) \h f_{\vec R'}^*(\vec k') \h V^{(C)}_{n_1 n_2 n_3 n_4}(\vec k, \h \vec s - \vec k, \h \vec s - \vec k', \h \vec k')  \label{doublek} \\[3pt]
 & = \frac 1 {N^2} \h \sum_{\vec k, \h \vec k'} \h \sum_{\vec R_1, \h \vec R_2} f_{\vec R}(\vec k) \h f_{\vec R'}^*(\vec k') \h f^*_{\vec R_1}(\vec k) \h f_{\vec R_2}(\vec k') \, C^{\vec R_1 \vec R_2}_{n_1 n_2 n_3 n_4}(\vec k + \vec k' - \vec s) \,. \label{zw_1}
\end{align}
We identify the form factors with plane-wave functions, $f_{\vec R}(\vec k) = \exp(-\textnormal i \vec k \cdot \vec R)$, and perform a Fourier transformation of the matrix $\mathbbmsl C$, i.e.,
\begin{align}
 \mathbbmsl C_{n_1 n_2 n_3 n_4}(\vec k) & = \sum_{\vec R} \e^{-\textnormal i \vec k \cdot \vec R} \,\h \mathbbmsl C_{n_1 n_2 n_3 n_4}(\vec R) \,, \\[5pt]
 \mathbbmsl C_{n_1 n_2 n_3 n_4}(\vec R) & = \frac 1 N \sum_{\vec k} \e^{\textnormal i \vec k \cdot \vec R} \,\h \mathbbmsl C_{n_1 n_2 n_3 n_4}(\vec k) \,.
\end{align}
Thus, we transform Eq.~\eqref{zw_1} into
\begin{equation}
\frac{1}{N^2} \h \sum_{\vec k, \h \vec k'} \h \sum_{\vec R_1, \h \vec R_2,\h  \vec R''} \e^{\textnormal i \vec k \cdot (-\vec R + \vec R_1 - \vec R'')} \, \e^{\textnormal i \vec k' \cdot (\vec R' - \vec R_2 - \vec R'')} \, \e^{\textnormal i\vec s \cdot \vec R''} \, C^{\vec R_1 \vec R_2}_{n_1 n_2 n_3 n_4}(\vec R'') \,.
\end{equation}
Next, evaluating the sums over $\vec k$ and $\vec k'$ yields two delta functions, which can be used 
to eliminate $\vec R_1$ and $\vec R_2$\h. Hence, we arrive at
\begin{equation} \label{proj_PD}
\hat P[V^{(C)}_{n_1 n_2 n_3 n_4}]_{\vec R \vec R'}(\vec s) = \sum_{\vec R''} \h \e^{\textnormal i \vec s \cdot \vec R''} \h C_{n_1 n_2 n_3 n_4}^{\vec R + \vec R''\mh, \h \vec R'-\vec R''}(\vec R'') \,.
\end{equation}
In particular, we note that this procedure (as used already in Ref.~\citep{Wang12}) has reduced the double-wavevector sums of Eq.~\eqref{doublek} to a single sum over form-factor indices $\vec R''$. This sum is usually finite because the truncated form-factor matrices $C_{n_1 n_2 n_3 n_4}^{\vec R_1, \h \vec R_2}(\vec R)$ are nonzero only for a finite number of components   $\vec R_1$ and $\vec R_2$.

Similarly, we find (using the property \eqref{antisym_cd} or \eqref{antisym}),
\begin{align}
 & \hat P[V^{(D)}_{n_1 n_2 n_3 n_4}]_{\vec R \vec R'}(\vec s) = -\e^{\textnormal i \vec s \cdot \vec R'} \h \sum_{\vec R''} \h \e^{\textnormal i\vec s \cdot \vec R''} \h C^{\vec R + \vec R''\mh, \h -\vec R' - \vec R''}_{n_1 n_2 n_4 n_3}(\vec R'') \,.
\end{align}
Next, we perform the projections in the $D$ channel. Again, we have
\begin{equation}
 \hat{\mathbbmsl C}[V_{n_1 n_2 n_3 n_4}] = \hat{\mathbbmsl C}[V^{(0)}_{n_1 n_2 n_3 n_4}] + \hat{\mathbbmsl C}[V^{(P)}_{n_1 n_2 n_3 n_4}] + \mathbbmsl C_{n_1 n_2 n_3 n_4} + \hat{\mathbbmsl C}[V^{(D)}_{n_1 n_2 n_3 n_4}] \,,
\end{equation}
where we have used Eq.~\eqref{def_mat_D}. A straightforward calculation gives
\begin{equation}
 \hat C[V^{(P)}_{n_1 n_2 n_3 n_4}]_{\vec R \vec R'}(\vec u) = \sum_{\vec R''} \h \e^{\textnormal i\vec u \cdot \vec R''}  P_{n_1 n_2 n_3 n_4}^{\vec R + \vec R''\mh, \h \vec R' - \vec R''}(\vec R'') \,,
\end{equation}
as well as
\begin{equation} \label{proj_DC}
 \hat C[V^{(D)}_{n_1 n_2 n_3 n_4}]_{\vec R \vec R'}(\vec u) = -\sum_{\vec R''} \h \e^{\textnormal i \vec u \cdot \vec R''} \h C^{\vec R - \vec R' + \vec R''\mh, \h \vec R''}_{n_1 n_2 n_4 n_3}(-\vec R') \,.
\end{equation}
In summary, in the non-SU(2)-symmetric case considered here, the channel-decomposed RGE for the matrices $\mathbbmsl P$ and $\mathbbmsl C$ are given by Eqs.~\eqref{rge_Pmat}, \eqref{rge_Dmat}, \eqref{loopminus}, \eqref{loopplus}, \eqref{sum_P}, \eqref{proj_PD}--\eqref{proj_DC}.

\section{Orbital versus band basis} \label{sec_orb_band}

\subsection{Transformation laws}

For multiband systems, the vertex function can be represented either in the {\itshape band basis} or in the {\itshape orbital basis}, where these terms refer to the respective single-particle bases. Given the unitary matrix $U_{\sigma n}(\vec k)$ which diagonalizes the single-particle Hamiltonian, these two representations of the vertex function are related as follows (see e.g.~\citep{Schober}):
\begin{align}
 & V_{n_1 n_2 n_3 n_4}(\vec k_1, \vec k_2, \vec k_3) = \label{trafo} \\[5pt] \nonumber
 & \sum_{\sigma_1, \ldots, \sigma_4} U_{\sigma_1 n_1}^*(\vec k_1) \h U_{\sigma_2 n_2}^*(\vec k_2) \h V_{\sigma_1 \sigma_2 \sigma_3 \sigma_4}(\vec k_1, \vec k_2, \vec k_3) \h U_{\sigma_3 n_3}(\vec k_3) \h U_{\sigma_4 n_4}(\vec k_1 + \vec k_2 - \vec k_3) \,.
\end{align}
As mentioned before, we denote band indices by Latin letters and orbital indices by Greek letters. Later, we will consider the Rashba model as a two-band model, where the $\sigma_i$ can be identified with {\itshape spin indices}. In any case, given these two representations of the vertex function, we can also define the respective channel projections  $\hat P[V_{\sigma_1 \sigma_2 \sigma_3 \sigma_4}]$ or $\hat P[V_{n_1 n_2 n_3 n_4}]$ as in \S\,\ref{sec_channel} (and similarly for $D$ and $C$). For deriving the transformation laws between these different matrices, we start from Eq.~\eqref{def_P} in the band basis, i.e.,
\begin{equation}
  \hat P[V_{n_1 n_2 n_3 n_4}]_{\vec R \vec R'}(\vec s) = \frac{1}{N^2} \sum_{\vec k, \h \vec k'} f_{\vec R}(\vec k) \h f^*_{\vec R'}(\vec k') \, V_{n_1 n_2 n_3 n_4}(\vec k, \h \vec s - \vec k, \h \vec s - \vec k') \,.
\end{equation}
We switch over to the spin basis using Eq.~\eqref{trafo}, and then employ the converse relation \eqref{def_P_inv} in the spin \linebreak basis, i.e., \smallskip
\begin{equation}
  V_{\sigma_1 \sigma_2 \sigma_3 \sigma_4}(\vec k, \h \vec s - \vec k, \h \vec s - \vec k') = \sum_{\vec R_1, \h \vec R_2} f_{\vec R_1}^*(\vec k) \h f_{\vec R_2}(\vec k') \, \hat P[V_{\sigma_1 \sigma_2 \sigma_3 \sigma_4}]_{\vec R_1 \vec R_2}(\vec s) \,. \smallskip
\end{equation}
Thus, we arrive at the following transformation law for the particle-particle projection:
\begin{equation} \label{trafo_P}
\begin{aligned}
 & \hat P[V_{n_1 n_2 n_3 n_4}]_{\vec R \vec R'}(\vec s) = \sum_{\sigma_1, \ldots, \sigma_4} \h \sum_{\vec R_1, \vec R_2} \bigg( \frac 1 N \sum_{\vec k} f_{\vec R}(\vec k) \h f_{\vec R_1}^*(\vec k) \h U^*_{\sigma_2 n_2}(\vec s - \vec k) \h U_{\sigma_1 n_1}^*(\vec k) \bigg) \\[5pt]
 & \times \hat P[V_{\sigma_1 \sigma_2 \sigma_3 \sigma_4}]_{\vec R_1 \vec R_2}(\vec s) \, \bigg( \frac 1 N \sum_{\vec k'} f_{\vec R_2}(\vec k') \h f_{\vec R'}^*(\vec k') \h U_{\sigma_3 n_3}(\vec s - \vec k') \h U_{\sigma_4 n_4}(\vec k') \bigg) \,.
\end{aligned}
\end{equation}
Defining the matrix
\begin{equation}
 U^{\vec R \vec R'}_{\sigma_1 \sigma_2 n_1 n_2}(\vec s) := \frac 1 N \sum_{\vec k} f_{\vec R}(\vec k) \h f_{\vec R'}^*(\vec k) \, U_{\sigma_1 n_1}(\vec s - \vec k) \h U_{\sigma_2 n_2}(\vec k) \,,
\end{equation}
we can write Eq.~\eqref{trafo_P} in matrix form as
\begin{equation} \label{mat_trafo_P}
 \hat{\mathbbmsl P}[V_{n_1 n_2 n_3 n_4}](\vec s) = \sum_{\sigma_1, \ldots, \sigma_4} [\mathbbmsl U_{\sigma_2 \sigma_1 n_2 n_1}(\vec s)]^{\dagger} \, \hat{\mathbbmsl P}[V_{\sigma_1 \sigma_2 \sigma_3 \sigma_4}](\vec s) \, \mathbbmsl U_{\sigma_3 \sigma_4 n_3 n_4}(\vec s) \,.
\end{equation}
The converse equation reads
\begin{equation}
 \hat{\mathbbmsl P}[V_{\sigma_1 \sigma_2 \sigma_3 \sigma_4}](\vec s) = \sum_{n_1, \ldots, n_4} \mathbbmsl U_{\sigma_2 \sigma_1 n_2 n_1}(\vec s) \, \hat{\mathbbmsl P}[V_{n_1 n_2 n_3 n_4}](\vec s) \, [\mathbbmsl U_{\sigma_3 \sigma_4 n_3 n_4}(\vec s)]^{\dagger} \,.
\end{equation}
Similarly, with the matrix
\begin{equation}
 X^{\vec R \vec R'}_{\sigma_1 \sigma_2 n_1 n_2}(\vec u) := \frac 1 N \sum_{\vec k} f_{\vec R}(\vec k) \h f_{\vec R'}^*(\vec k) \, U_{\sigma_1 n_1}(\vec k) \h U^*_{\sigma_2 n_2}(\vec k - \vec u) \,,
\end{equation}
we obtain the transformation laws
\begin{align} \label{mat_trafo_D}
 \hat{\mathbbmsl C}[V_{n_1 n_2 n_3 n_4}](\vec u) & = \sum_{\sigma_1, \ldots, \sigma_4} [\mathbbmsl X_{\sigma_1 \sigma_3 n_1 n_3}(\vec u)]^\dagger \, \hat{\mathbbmsl C}[V_{\sigma_1 \sigma_2 \sigma_3 \sigma_4}](\vec u) \, \mathbbmsl X_{\sigma_4 \sigma_2 n_4 n_2}(\vec u) \,, \\[5pt] \hat{\mathbbmsl D}[V_{n_1 n_2 n_3 n_4}](\vec t) & = \sum_{\sigma_1, \ldots, \sigma_4} [\mathbbmsl X_{\sigma_1 \sigma_4 n_1 n_4}(\vec t)]^\dagger \, \hat{\mathbbmsl D}[V_{\sigma_1 \sigma_2 \sigma_3 \sigma_4}](\vec t) \, \mathbbmsl X_{\sigma_3 \sigma_2 n_3 n_2}(\vec t) \,.  \label{mat_trafo_C}
\end{align}
These transformation laws can be used to switch between the band basis and the orbital/spin basis in the TU-fRG scheme.

\subsection{RGE in orbital basis}

While in \S\,\ref{subsec_deriv} we have derived the channel-decomposed RGE in the band basis, one can analogously deduce the RGE in the orbital basis (see also \citep[\S\,III.E]{Schober}). For example, the exchange propagator in the orbital basis,
\begin{equation}
 P^{\vec R \vec R'}_{\sigma_1 \sigma_2 \sigma_3 \sigma_4}(\vec s) := \hat P[V^{(P)}_{\sigma_1 \sigma_2 \sigma_3 \sigma_4}]_{\vec R \vec R'}(\vec s) \,,
\end{equation}
fulfills the following RGE:
\begin{equation} \label{RGE_spin}
 \dot{\mathbbmsl P}_{\sigma_1 \sigma_2 \sigma_3 \sigma_4}(\vec s) = \sum_{\tau_1, \ldots, \tau_4} \hat{\mathbbmsl P}[V_{\sigma_1 \sigma_2 \tau_1 \tau_2}](\vec s) \, \mathbbmsl L^-_{\tau_1 \tau_2 \tau_3 \tau_4}(\vec s) \, \hat{\mathbbmsl P}[V_{\tau_4 \tau_3 \sigma_3 \sigma_4}](\vec s) \,,
\end{equation}
where the particle-particle loop in the orbital basis depends on {\itshape four} orbital indices and is given in terms of its counterpart in the band basis, Eq.~\eqref{loopminus}, by
\begin{equation} \label{loop_orbital}
 \mathbbmsl L^-_{\tau_1 \tau_2 \tau_3 \tau_4}(\vec s) = \sum_{\ell_1, \ell_2} \mathbbmsl U_{\tau_1 \tau_2 \ell_1 \ell_2}(\vec s) \, \mathbbmsl L^-_{\ell_1 \ell_2}(\vec s) \, [\mathbbmsl U_{\tau_3 \tau_4 \ell_1 \ell_2}(\vec s)]^\dagger \,.
\end{equation}
Similarly, one can derive the RGE for the matrices $\mathbbmsl C$ and $\mathbbmsl D$ in the orbital basis.

\subsection{Discussion} \label{sec_compare}

We conclude this section by a short comparison between the band basis and the orbital basis, focusing on practical aspects of the numerical implementation. The question which picture to choose arises in systems with more than one site per sublattice in the case of spin-rotational invariance, and even with one site per sublattice if spin-orbit coupling makes the single-particle Hamiltonian non-diagonal in the spins. While SM-fRG works like \citep{Wang12} use the orbital basis, previous TU-fRG studies like \citep{David17} and \citep{David17Coul} were fomulated in the band basis.

On the one hand, the band basis diagonalizes the free Hamiltonian and the free Green function, and therefore the loop terms depend on only two band indices (see the explicit expressions \eqref{def_lmp}--\eqref{disc_2b}). By contrast, in the orbital basis the Green function depends on two orbital indices, and correspondingly the loop terms depend on four orbital indices (see Eq.~\eqref{loop_orbital}). As a consequence, the CPU time required for a numerical implementation of the TU-fRG scales considerably with the number of orbitals $n_{\rm o}$ or the number of bands $n_{\rm b}$. 
Concretely, taking $n_{\rm b}= n_{\rm o}$, if we wish to compute the whole vertex function in the orbital basis, e.g.~via Eq.~\eqref{RGE_spin}, then we need to (i) calculate all $n_{\rm o}^4$ entries of the exchange propagator matrices, (ii) perform $n_{\rm o}^2$ summations for calculating the loop terms via Eq.~\eqref{loop_orbital}, and (iii) perform the $n_{\rm o}^4$ summations on the right-hand side of the RGE \eqref{RGE_spin}. Thus, the CPU time scales as $n_{\rm o}^{10}$, i.e., with the 10th power of the number of orbitals. By contrast, since in the band basis the loop terms are given explicitly by Eqs.~\eqref{def_lmp}--\eqref{disc_2b}, and since the right-hand sides of the RGE \eqref{rge_Pmat}--\eqref{rge_Cmat} require only a summation over two band indices, the CPU time scales only as $n_{\rm b}^6$, i.e., with the 6th power of the number of bands.

On the other hand, the initial two-particle vertex is usually given in the orbital basis, where it often has a weak momentum dependence (corresponding to a local interaction in real space). By transforming the vertex into the band basis via Eq.~\eqref{trafo}, it will typically acquire a complicated structure in momentum space (corresponding to a potentially complicated non-local form in real space). For the exchange propagators of the TU-fRG, this implies that a form-factor expansion in the orbital basis will lead to much faster convergence than in the band basis. In fact, we will confirm this last observation in the next section, where we will derive analytical expressions for the particle-particle exchange propagator in the Rashba model, both in the orbital basis and in the band basis (see Eqs.~\eqref{sol_spin} and \eqref{Pinband}, respectively). It should also be noted that e.g. \citep{David17Coul} discusses explicitly the convergence issues entailed by the projection of a longer-ranged initial interaction. Furthermore, at least the dominating parts of the bare interaction can usually be understood as a sum of terms with pairwise identical orbital indices. Hence, it may constitute a useful approximation to focus on the renormalization of these terms, which improves the scaling with $n_{\rm o}$.  

\section{Application to Rashba model} \label{sec_appl}

We consider a general two-band model with the single-particle Hamiltonian given by
\begin{equation} \label{gen_Ham}
 H_0(\vec k) = f(\vec k) \h \mathbbm 1 + \vec g(\vec k) \cdot \vec \sigma \,,
\end{equation}
where $f(\vec k)$ and $g_i(\vec k)$ ($i = x, y, z$) are real functions, and $\vec \sigma = (\sigma_x, \sigma_y, \sigma_z)$ denotes the vector of Pauli matrices. We assume time-reversal symmetry, which implies that $f(\vec k) = f(-\vec k)$ as well as $\vec g(\vec k) = -\vec g(-\vec k)$ (see e.g.~\citep[Appendix A.2]{Schober}). In particular, 
the Rashba model is recovered from the more general Eq.~\eqref{gen_Ham} by setting
\begin{equation}
f(\vec k) = \hbar^2 |\vec k|^2 / 2 m^* \,, \quad g_x(\vec k) = -k_y \,, \quad g_y(\vec k) = k_x \,, \quad g_z(\vec k) = 0 \,.
\end{equation}
For this model with a local initial interaction, it turned out \citep{Schober} that the particle-hole terms are negligible in the RG flow. The remaining particle-particle ladder could be resummed analytically in the spin basis \citep[\S\,III.E]{Schober}. Correspondingly, we here neglect the matrix $\mathbbmsl C$ and restrict attention to the remaining RGE \eqref{RGE_spin} for the matrix
\begin{equation}
 \mathbbmsl P_{\sigma_1 \sigma_2 \sigma_3 \sigma_4}(\vec s) \h = \h \hat{\mathbbmsl P}[V^{(P)}_{\sigma_1 \sigma_2 \sigma_3 \sigma_4}](\vec s) \h = \h \hat{\mathbbmsl P}[V_{\sigma_1 \sigma_2 \sigma_3 \sigma_4}](\vec s) - \hat{\mathbbmsl P}[V^{(0)}_{\sigma_1 \sigma_2 \sigma_3 \sigma_4}](\vec s) \,.
\end{equation}
For an initial onsite interaction,
\begin{equation}
 V^{(0)}_{\sigma_1 \sigma_2 \sigma_3 \sigma_4}(\vec k_1, \vec k_2, \vec k_3) = \frac U 2 \h (\delta_{\sigma_1 \sigma_3} \h  \delta_{\sigma_2 \sigma_4} - \delta_{\sigma_1 \sigma_4} \h \delta_{\sigma_2 \sigma_3} ) \,,
\end{equation}
the analytical solution of this RGE reads as follows:
\begin{equation} \label{sol_spin}
 P^{\vec R \vec R'}_{\sigma_1 \sigma_2 \sigma_3 \sigma_4}(\vec s) = -\frac 1 2 \h \big(\hh g_\Lambda(\vec s) + U\hh \big) \, (\delta_{\sigma_1 \sigma_3} \h  \delta_{\sigma_2 \sigma_4} - \delta_{\sigma_1 \sigma_4} \h \delta_{\sigma_2 \sigma_3} ) \, \delta_{\vec R, \hh \vec 0} \, \delta_{\vec R'\mh, \hh \vec 0} \,,
\end{equation}
with the scale-dependent scalar function
\begin{equation}
g_\Lambda(\vec s) = -U \h \bigg( 1 + U \int_{\Lambda_0}^{\Lambda} B_{\Lambda'}(\vec s) \h \de \Lambda' \bigg)^{\!\!-1} \,.
\end{equation}
Here, we have defined the auxiliary function
\begin{align}
 B(\vec s) & = \frac 1 2 \sum_{\tau_1, \ldots, \tau_4} (\delta_{\tau_1 \tau_3} \h \delta_{\tau_2 \tau_4} - \delta_{\tau_1 \tau_4} \h \delta_{\tau_2 \tau_3} ) \, [\mathbbmsl L^-_{\tau_1 \tau_2 \tau_3 \tau_4}(\vec s)]_{\vec 0 \vec 0} \\[5pt]
 & = \frac 1 2 \sum_{\tau_1, \ldots, \tau_4} (\delta_{\tau_1 \tau_3} \h \delta_{\tau_2 \tau_4} - \delta_{\tau_1 \tau_4} \h \delta_{\tau_2 \tau_3} ) \, \frac 1 N \sum_{\vec k} L^-_{\tau_1 \tau_2 \tau_3 \tau_4}(\vec s - \vec k, \h \vec k) \,,
\end{align}
which coincides with \citep[Eq.~(178)]{Schober}. Remarkably, the exchange propagator \eqref{sol_spin} in the spin basis remains local at any scale below the initial scale $\Lambda_0$.

Next, we transform Eq.~\eqref{sol_spin} into the band basis by means of Eq.~\eqref{mat_trafo_P}. Taking advantage of the formula (40) in Ref.~\citep{Schober}, we arrive at
\begin{align} \label{Pinband}
 P_{n_1 n_2 n_3 n_4}^{\vec R \vec R'}(\vec s) & = -\frac 1 8 \h \big(\hh g(\vec s) + U\hh\big) \h \frac 1 {N^2} \sum_{\vec k, \h \vec k'}  f_{\vec R}(\vec k) \h f^*_{\vec R'}(\vec k') \h  \Big[ \hh n_1 n_3 \, \e^{\mathrm i \varphi(\vec s - \vec k') - \mathrm i \varphi(\vec k)} \\[5pt] \nonumber 
 & \quad \, + n_2 \h n_4 \, \e^{\mathrm i \varphi(\vec k') - \mathrm i \varphi(\vec s - \vec k)} - n_1 n_4 \, \e^{\mathrm i \varphi(\vec k') - \mathrm i \varphi(\vec k)} - n_2 \h n_3 \, \e^{\mathrm i \varphi(\vec s - \vec k') - \mathrm i \varphi(\vec s - \vec k)} \Big] \,,
\end{align}
where $\varphi(\vec k)$ denotes the polar angle of the vector $\vec g(\vec k)$, and where the band indices $n_i \in \{-, +\}$ label the lower or upper band, respectively (see \citep[\S\,II]{Schober}). Further using Eq.~\eqref{def_P}, we 
read off the vertex function in the band basis as
\begin{align} \label{vert_band}
 V_{n_1\ldots n_4}(\vec k, \h \vec s - \vec k, \h \vec s - \vec k') & = -\frac 1 8 \h\hh g(\vec s) \h  \Big[ \hh n_1 n_3 \, \e^{\mathrm i \varphi(\vec s - \vec k') - \mathrm i \varphi(\vec k)} + n_2 \h n_4 \, \e^{\mathrm i \varphi(\vec k') - \mathrm i \varphi(\vec s - \vec k)} \\[5pt] \nonumber 
 & \hspace{2.2cm} - n_1 n_4 \, \e^{\mathrm i \varphi(\vec k') - \mathrm i \varphi(\vec k)} - n_2 \h n_3 \, \e^{\mathrm i \varphi(\vec s - \vec k') - \mathrm i \varphi(\vec s - \vec k)} \Big] \,.
\end{align}
In particular, for $\vec s = \vec 0$, this reduces to
\begin{equation}
 V_{n_1 \ldots n_4}(\vec k, \h -\vec k, \h -\vec k') = \frac 1 2 \h\hh g(\vec 0) \, \delta_{n_1 n_2} \h \delta_{n_3 n_4} \h n_1 \hh n_4 \, \e^{\mathrm i \varphi(\vec k') - \mathrm i \varphi(\vec k)} \,,
\end{equation}
which is again consistent with the results of Ref.~\citep{Schober}. We note that the exchange propagator in the band basis, Eq.~\eqref{Pinband}, is not local anymore (in contrast to the expression \eqref{sol_spin} in the orbital basis). Thus, in accordance with the remarks in \S\,\ref{sec_compare}, a numerical implementation of the TU-fRG in the band basis would require one to keep track of a large number of form factors. 

Finally, the above analytical solution for the Rashba model also gives some insights into the general advantages of the TU-fRG technique. In fact, our expression \eqref{vert_band} in the band basis shows a complicated momentum dependence, where for $\vec s \not = \vec 0$ the vertex function does not only depend on the angular variables $\varphi(\vec k)$ and $\varphi(\vec k')$. For reproducing the correct form of the effective interaction in a numerical implementation, it would therefore be necessary to take into account a sufficiently fine mesh of discrete wavevectors over the whole Brillouin zone. This, however, would be computationally demanding for an ordinary Fermi surface patching scheme (where the CPU time scales with the fourth power of the number of patches). By contrast, the TU-fRG scales only linearly with the number of Bloch momenta \citep{Julian17} and therefore allows for a much higher resolution of the momentum dependencies.

\section{Conclusion and Outlook}

We have reviewed the TU-fRG as a flexible and unbiased tool for investigating correlated electron systems, and we have adapted it for application to multiband systems with spin-orbit coupling. In particular, we have defined the single-channel coupling functions and  exchange propagators in the general case without SU(2) symmetry. As a consequence of an antisymmetry relation, only two exchange propagators (which correspond to the particle-particle and the crossed particle-hole term) actually need to be considered. Furthermore, we have derived the corresponding flow equations, which are of a particularly simple form (see Eqs.~\eqref{rge_Pmat}--\eqref{rge_Cmat}, and compare them to the corresponding equations in the SU(2)-symmetric case, i.e., \citep[Eqs.~(22)--(24)]{Julian17}). In fact, these flow equations are analogous to the corresponding equations in the SM-fRG \citep{Wang4}. On the right-hand side of the flow equations, projections between the different channels have to be performed, for which we have derived explicit expressions in \S\,\ref{sec_proj}. We have also compared the different formulations of the TU-fRG in the band basis and the orbital basis. Finally, we have analytically solved the channel-decomposed RGE in the particle-particle channel for the Rashba model with a local interaction, whereby we have shown the consistency of this solution with the ladder resummation of Ref.~\citep{Schober}.

To put this work into perspective, let us summarize the main advantages of the TU-fRG and outline possible future applications: First, in a numerical implementation the CPU time scales only linearly with the number of discrete Bloch momenta, which allows one to reach an extremely high momentum-space resolution. At the same time, the time-consuming part of calculating the right-hand sides of the flow equations can be parallelized efficiently on a large number of compute nodes \citep{Julian17}. For these reasons, the TU-fRG may be particularly advantageous in cases where Fermi surface patching with an insufficient momentum resolution influences the leading instability or the form of the effective interaction (see e.g.~\citep{Volpez16}). Furthermore, the speed-up gained from the efficient parallelization can be used to treat complicated multiband systems (for a proof of principle see \citep{David17}) or long-range interactions, which generally lead to sharp structures in momentum space (see \citep{David17Coul}). Other possible future directions include the treatment of three-dimensional band structures (where usually, the implementation of Fermi surface patching is numerically too expensive), the investigation of frequency-dependent interaction vertices, or the consideration of self-energy feedback onto the flow of the two-particle vertex. As general advantages of the TU-fRG, we further mention its applicability in wide parameter ranges (in which it complements non-perturbative methods such as lattice QMC), and its unbiasedness with regard to different (and possibly competing) ordering tendencies \citep{Metzner}.

With the present extension of the TU-fRG to non-SU(2)-symmetric systems, we have further enlarged its range of applications to embrace the important class of spin-orbit coupled materials. These include non-centrosymmetric (and possibly topological) superconductors \citep{Smidman17}, Rashba semiconductors \citep{Ishizaka, Liebmann16}, and Weyl semimetals \citep{Zyuzdin12}. In particular, the three-dimensional dispersion of Weyl semimetals has hindered so far a direct application of the fRG with Fermi surface patching, whereas their investigation using TU-fRG is feasible and currently underway. Thus, we expect the TU-fRG and its generalization presented here to play an important role in the quantitative description of correlated quantum materials.

\section*{Funding}

This research was supported by the DFG grants HO 2422/10-1, 11-1, and 12-1 and by the DFG RTG 1995.

\section*{Acknowledgments}

We thank Christian J.~Eckhardt, Julian Lichtenstein, Manfred Salmhofer, David S\'{a}nchez de la Pe\~na, Michael M.~Scherer and Qianghua Wang for discussions.

\bibliographystyle{frontiersinHLTH&FPHY} % for Health, Physics and Mathematics articles
\bibliography{mybib}

%%% Make sure to upload the bib file along with the tex file and PDF
%%% Please see the test.bib file for some examples of references

\end{document}